\documentclass[twocolumn,showpacs,preprintnumbers,amsmath,amssymb]{revtex4}
\usepackage{graphicx}% Include figure files
\usepackage{dcolumn}% Align table columns on decimal point
\usepackage{bm}% bold math

\begin{document}

\title{Magnetic Black Holes Are Also Unstable}

\author{Sang Pyo Kim}
\email{sangkim@kunsan.ac.kr}
\affiliation{Department of Physics,
   Kunsan National University,
   Kunsan 573-701, Korea}

\author{Don N. Page}
\email{don@phys.ualberta.ca}
\affiliation{Theoretical Physics Institute, Department of Physics,
   University of Alberta,
   Edmonton, Alberta, Canada T6G 2J1}

\date{2004 Feb. 24}

\begin{abstract}

Most black holes are known to be unstable to emitting Hawking
radiation (in asymptotically flat spacetime).  If the black holes are
non-extreme, they have positive temperature and emit thermally.  If
they are extremal rotating black holes, they still spontaneously emit
particles like gravitons and photons.  If they are extremal
electrically charged black holes, they are unstable to emitting
electrons or positrons.  The only exception would be extreme
magnetically charged black holes if there do not exist any magnetic
monopoles for them to emit.  However, here we show that even in this
case, vacuum polarization causes all magnetic black holes to be
unstable to emitting smaller magnetic black holes.

 \end{abstract}

\maketitle

\section{Introduction}

Hawking \cite{H1,H2} showed that black holes in asymptotically flat
spacetime are unstable to emitting thermal radiation at temperature
\begin{eqnarray}
T = {\kappa\over 2\pi}
  = {\sqrt{M^2-Q^2-J^2/M^2}\over
  2\pi(2M^2-Q^2+2M\sqrt{M^2-Q^2-J^2/M^2})},
\label{1}
\end{eqnarray}
where $\kappa$ is the horizon surface gravity, $M$ is the mass, $Q$ is
the charge, and $J$ is the angular momentum, using Planck units in
which $\hbar = c = G = k_{\rm Boltzmann} = 1/(4\pi\epsilon_0) = 1$.

This Hawking radiation decreases the mass and angular momentum of the
hole at the following rate:
\begin{eqnarray}
-{d\over dt} {M \choose J} &=&
\sum_{j,l,m,p}\int_{\mu_j}^{\infty}{d\omega\over 2\pi}
 {\omega \choose J} \nonumber\\
&& \times
 {\Gamma_{jlmp}(\omega)\over \exp{[(\omega-e\Phi-m\Omega)/T]}\mp 1}.
\label{2}
\end{eqnarray}
Here the sum is over the species (labeled by $j$), the total angular
momentum number $l$ of each wave mode, the axial angular momentum number $m$
of the mode, and the polarization $p$ of the mode, and the integral is
over the frequency $\omega$ of the mode, from the rest mass $\mu_j$ of
the species to infinity.

In the exponential in the denominator, $e$ represents the charge of
the particle being emitted, $\Phi$ is the electrostatic potential of
the hole, and $\Omega$ is the angular velocity of the hole.  The upper
sign after the exponential ($-$) is for bosons, and the lower sign ($+$)
is for fermions.  The absorption coefficient is
\begin{equation}
\Gamma_{jlmp}(\omega) = 1 - A_{jlmp}(\omega),
\label{3}
\end{equation}
where $A_{jlmp}(\omega)$ is the classical amplification coefficient
for the mode, the ratio of the outgoing to the ingoing flux at spatial
infinity for the mode with the boundary condition of ingoing group
velocity at the black hole horizon.  (The absorption coefficient is
nonnegative except for bosonic superradiant modes that have a negative
exponent in the exponential so that the denominator is negative, and
then the absorption coefficient is negative so that the Hawking
radiation in each mode drains energy from the hole.)

Now we can see that there are various special cases.

(1)  A Schwarzschild black hole, with $Q = J = \Phi = \Omega = 0$,
emits thermal radiation with temperature $T = 1/(2\pi M)$.

(2)  An uncharged extreme rotating Kerr black hole, with $Q = \Phi =
0$ but $J^2 = M^4$, has $T = 0$ but emits at a nonzero rate in the
energy range $\mu_j < \omega < m\Omega$ where the exponent of the
exponential is $-\infty$:
\begin{eqnarray}
-{d\over dt} {M \choose J} =
\sum_{j,l,m,p}\int_{\mu_j}^{m\Omega}{d\omega\over 2\pi}
 {\omega \choose m}
 \left[\mp\Gamma_{jlmp}(\omega)\right].
\label{4}
\end{eqnarray}
For bosons, this is the spontaneous emission corresponding to the
stimulated emission that is a quantum description of superradiant
amplification, $A_{jlmp}(\omega) > 1$.  For fermions, there is an
analogous spontaneous emission even though the Pauli exclusion
principle prevents the amplification factor from being greater than
unity (as Richard Feynman explained to William Press, Saul Teukolsky,
and one of the authors (D.N.P.) around 1972 while drawing diagrams
on a blackboard and saying,
``I'm supposed to be good at these diagrams'').  Therefore, even
though there is no true superradiance for fermions, one can say that
there is a ``superradiant'' range for each where $\omega-e\Phi-m\Omega
< 0$ and hence where $\exp{[(\omega-e\Phi-m\Omega)/T]} = 0$ for $T = 0$.

(3)  An extreme charged nonrotating Reissner-Nordstrom black hole,
with $Q = M$, $\Phi = 1$, and $J = \Omega = 0$, also has $T = 0$ but
emits charged particles of the same sign of charge as the hole (say
positive for concreteness) at the rate
\begin{eqnarray}
-{d\over dt}{M  \choose J} =
\sum_{j,l,m,p}\int_{\mu_j}^{e\Phi}{d\omega\over 2\pi}
 {\omega \choose m}
 \left[\mp\Gamma_{jlmp}(\omega)\right].
\label{5}
\end{eqnarray}

Henceforth we shall have little use for the axial angular momentum, so
let us instead use $m$ for the rest mass $\mu_j$ of the species of
particles being emitted.  For an extreme Reissner-Nordstrom black hole
of mass $M$ and charge $Q = M$, particles of mass $m$ and charge $e$
are emitted within the ``superradiant'' range
\begin{equation}
m < \omega < e\Phi = {eQ\over r_{+}}
  = {eQ\over M+\sqrt{M^2-Q^2}} = e = \sqrt{\alpha},
\label{6}
\end{equation}
where the last equality applies for an elementary particle with the
charge of the positron, which in Planck units is the square root of
the electromagnetic fine structure constant $\alpha \approx
1/137.036$.

Thus extreme Reissner-Nordstrom black holes can emit particles with
$e/m > 1$.  This is true for all known elementary charged particles,
with the positron having $e/m \approx 2.04 \times 10^{21}$, more
than 21 orders of magnitude larger than unity.

In summary, it is known that all black holes that are neutral or have
just ordinary electric charge are unstable to losing mass by Hawking
emission.  However, we must examine the case of extremal black holes
with magnetic charge.

\section{Emission of ordinary magnetic magnetic monopoles}

If magnetic monopoles with magnetic charge
\begin{equation}
g = {n \over 2e}
\label{7}
\end{equation}
(the Dirac quantization condition with integer $n$) and mass $m < g$
exist, they can be emitted from an extreme magnetic black hole with
magnetic charge $P = M$, using $P$ to denote the magnetic charge of
the black hole in Planck units.

GUTs generally predict magnetic monopoles with masses
\begin{equation}
m \sim {M_G\over e^2} < g \sim {1 \over e},
\label{8}
\end{equation}
since the GUT unification scale $M_G$ generally is significantly more
than an order of magnitude lower than the Planck mass $M_{\rm Pl}
\equiv 1$, $M_G < e = \sqrt{\alpha} \sim 0.1 = 0.1 M_{\rm Pl}$.  If
these magnetic monopoles exist, extreme magnetic black holes would be
unstable to emitting these magnetic monopoles with magnetic charge $g$
greater than their mass $m$.

But what if there are no GUT monopoles?
If there are no GUT monopoles, one might think that extreme magnetic
black holes would be stable.  Classically, they would have magnetic
charge
\begin{equation}
P = {N \over 2e}
\label{9}
\end{equation}
and mass $M=P$.  No kinetic energy would be released if a large $P=M$
hole split into smaller holes with $\sum M_i = \sum P_i = P$.  Thus,
there would be no phase space available for this putative decay.

If the large extreme hole did split into smaller extreme holes, these
smaller holes (when not moving relative to each other, which would be
the case since there is no kinetic energy released) would have no
forces between them, since the attractive gravitational forces would
be precisely balanced by the repulsive magnetic forces:
\begin{equation}
-F_{\rm grav} = {M_1 M_2 \over r^2}
= F_{\rm mag} = {P_1 P_2 \over r^2}.
\label{10}
\end{equation}

\section{Origin of the quantum instability of a magnetic black hole}

{\em Vacuum polarization gives $M<P$ and makes extreme magnetic black
holes unstable to splitting}.

That is, the mass-to-charge ratio of an extremal magnetic charged
black hole is reduced below unity by vacuum polarization:
\begin{equation}
\mathcal{E}(P) \equiv {M_{\rm extreme}\over P} = 1 - \delta(P) < 1,
\label{11}
\end{equation}
where vacuum polarization gives a positive $\delta(P)$ that increases
with the magnetic field strength at the horizon,
\begin{equation}
B_{+} = {P \over r_{+}^2} \approx {1 \over P}.
\label{12}
\end{equation}
Hence smaller holes, with smaller $P$, have bigger $B_{+}$, bigger
$\delta(P)$, and smaller $\mathcal{E}(P)$.  Thus kinetic energy is
released when a large extreme black hole splits into smaller ones.

One might object that since the entropy of an extreme black hole is
\begin{equation}
S = {1\over 4}A = \pi r_{+}^2 \approx \pi P^2,
\label{13}
\end{equation}
entropy would be reduced when a large extremal black hole with
magnetic charge $P$ splits up into smaller extremal black holes with
$\sum P_i = P$.  However, in asymptotically flat spacetime with the
positive kinetic energy released by the splitting, there is an
infinite volume of phase space available and hence an infinite
capacity for entropy in the form of the positions and momenta of the
final black holes.  Thus there is no restriction from the second law
of thermodynamics against a large black hole splitting into smaller
ones in asymptotically flat spacetime, though there would be for a
space of finite volume or for a space of effectively finite volume,
such as anti-deSitter spacetime (a problem which shall be left for the
future).

\section{mass-to-charge ratio of extremal black
  holes}

The mass-to-charge ratio of extremal black holes, $\mathcal{E}(P)$
given by Eq. (\ref{11}), is shifted below unity by the effects of the
vacuum polarization of charged particle fields around the
magnetically charged black hole.

The largest effect will be by the charged particle field with the
lowest mass, the electron-positron field.  Therefore, consider the
one-loop effect of the electron-positron field on the vacuum.  This is
given by the 1936 Euler-Heisenberg Lagrangian \cite{EH} and dominates
for weak fields.

Define the quantity
\begin{equation}
b \equiv {eB \over m^2}
\label{14}
\end{equation}
which is dimensionless even without setting $G = 1$, where $m$ and $e$
are the mass and charge of the positron.  Then for a uniform magnetic
field $B$, the Lagrangian density through one loop in the
electron-positron field is
\begin{eqnarray}
L &=& -{B^2\over 8\pi}[1-{e^2\over\pi}I(b)],
\label{15} \\
I(b) &=& \int_0^\infty dx F(x) \exp{\left(-{x\over b}\right)},
\label{16} \\
F(x) &=& {1\over x^3}(1 + {1\over 3}x^2 - x \coth x).
\label{17}
\end{eqnarray}

Now we need to ask the question of when the magnetic field is
sufficiently homogeneous that the equations above for the Lagrangian
density of a uniform field are a good approximation for the Lagrangian
density of a non-uniform field.  For this, we can use the criteria
given by V. I. Ritus \cite{Ritus}, for a length scale $\lambda$ of the
inhomogeneity that for a magnetic black hole is the radius $r$:
\begin{eqnarray}
\lambda &=& r \gg \min\left({m\over eB},{1\over\sqrt{eB}}\right)
\nonumber\\
&=& \min\left({mr^2\over eP},{r\over\sqrt{eP}}\right)
        = \min\left({2mr^2\over N},{\sqrt{2}r\over\sqrt{N}}\right),
\label{18}
\end{eqnarray}
where Eq. (\ref{9}) gives $N = 2eP$, an integer.  That is, the
formulas for the Lagrangian density of a uniform magnetic field give a
good approximation for the Lagrangian density of the non-uniform field
outside an extreme magnetic black hole for
\begin{equation}
N \gg \min(mr,1).
\label{19}
\end{equation}

At the horizon, $r = r_{+} \approx P = N/(2e)$, so
\begin{equation}
mr_{+} \approx {m\over 2e}N \approx 2.45 \times 10^{-22} N \ll N,
\label{20}
\end{equation}
and so $N \gg \min(mr_{+},1)$ is always satisfied for extreme magnetic
black holes that are much larger than the Planck size and hence have
$N \gg 1$.

$N \gg \min(mr,1)$ will not be satisfied for $r > (2e/m) r_{+} \sim 4
\times 10^{21} r_{+}$, but at this radius the Lagrangian density will
be smaller than at the horizon by a factor of $\sim m^2/(2e)^2 \sim 6
\times 10^{-44}$, so that the effects of larger radii, where the
formulas above are not a good approximation, will be negligible.  That
is, since the dominant effect of the vacuum polarization is fairly
near the horizon, for large extreme magnetic black holes, we can
always neglect the (radial) inhomogeneity of the magnetic field.

Now we must calculate the black hole metric with vacuum polarization,
using the Euler-Heisenberg Lagrangian (\ref{15}).
The electromagnetic field tensor has the form
\begin{equation}
\mathbf{F} = P \sin\theta d\theta \wedge d\phi
 = B \widehat{d\theta} \wedge \widehat{d\phi},
\label{22}
\end{equation}
where $\widehat{d\theta}=d\theta/r$ and $\widehat{d\phi}=\sin\theta
d\phi/r$ are the orthonormal one-forms, and where the orthonormal
magnetic field strength is
\begin{equation}
B = {P \over r^2}.
\label{23}
\end{equation}
Here $r$ is a Schwarzschildean radial coordinate, defined so that
surfaces of constant $r$ are two-spheres with area $4\pi r^2$.

The vacuum polarization of electrically charged fields, such as the
electron-positron field, does not produce any density of magnetic
charge and so does not affect these formulas for the magnetic field,
which arise simply from the Maxwell equation $d\mathbf{F} = 0$ in the
absence of magnetic monopoles.  We are indeed assuming no magnetic
monopoles present in the theory (or else they would themselves make
the extreme black hole unstable, assuming that they have $m < g$),
other than black holes.

Now the magnetic field outside a large extreme magnetic black hole
would indeed produce a vacuum polarization of smaller magnetic black
holes so that $d\mathbf{F}$ would not quite be zero, or $Br^2$ would
not quite be constant.  However, this would be a very tiny effect.
Since the mass of these smaller magnetic black holes is so much larger
than the mass of electrons and positrons, the effects of their
vacuum polarization would be much smaller than that of the
electron-positron field.  Thus here we shall take $B = P/r^2$ as
essentially exact.

With the Euler-Heisenberg Lagrangian density $L$ given above, the
stress-energy tensor of the magnetic field and vacuum polarization is
\begin{equation}
T_{\alpha\beta}
= L g_{\alpha\beta} - 2 {dL\over d(B^2)}g^{\mu\nu}F_{\alpha\mu}F_{\beta\nu}.
\label{24}
\end{equation}
For a static spherically symmetric metric, this gives the orthonormal
components of the stress-energy tensor as
\begin{eqnarray}
\rho &\equiv& T_{\hat{0}\hat{0}} = -L
 = {B^2\over 8\pi}\left[1-{e^2\over\pi}I(b)\right], \\
\label{25}
P_r &\equiv& T_{\hat{r}\hat{r}} = + L
 = -{B^2\over 8\pi}\left[1-{e^2\over\pi}I(b)\right],
\label{26}
\end{eqnarray}
\begin{eqnarray}
P_{\perp} &\equiv& T_{\hat{\theta}\hat{\theta}} = L - 2B^2{dL\over d(B^2)}
 \nonumber\\
&=& {B^2\over 8\pi}\left[1-{e^2\over\pi}\left(I+b{dI\over db}\right)\right].
\label{27}
\end{eqnarray}

If $f' \equiv df/dr$, the static spherically symmetric metric
\begin{eqnarray}
ds^2 = &-&e^{2\psi(r)}\left(1-{2m(r)\over r}\right)dt^2
+\left(1-{2m(r)\over r}\right)^{-1}dr^2 \nonumber \\
&+& r^2 (d\theta^2 + \sin^2\!\theta\, d\phi^2)
\label{28}
\end{eqnarray}
has the Einstein field equations
\begin{equation}
\psi' = {4\pi r\over 1-2m/r}(\rho + P_r) = 0
\label{29}
\end{equation}
here, so $\psi = 0$ with a suitable choice of the time coordinate $t$,
and
\begin{eqnarray}
m' &=& 4\pi r^2 \rho = {1\over 2}r^2 B^2
 \left[1-{e^2\over\pi}I(b)\right] \nonumber\\
  &=& {P^2\over 2r^2}\left[1-{e^2\over\pi}I\left({eP\over m^2 r^2}\right)\right].
\label{30}
\end{eqnarray}

As a check on these equations, we can see that if we set $e^2/\pi =
\alpha/\pi = 0$ to ignore the vacuum polarization, we would get the
classical magnetic Reissner-Nordstrom metric with
\begin{equation}
m(r) = M - {P^2\over 2r},
\label{31}
\end{equation}
\begin{equation}
V = -g_{00} = g^{rr} \equiv 1 - {2m(r)\over r}
 = 1 - {2M\over r} + {P^2\over r^2}.
\label{32}
\end{equation}

To solve for the metric and mass $M = m(r=\infty)$, it is convenient
to define a critical magnetic charge
\begin{eqnarray}
P_{\ast} &\equiv& {N_{\ast}\over 2e} = {e\over m^2}
 \approx 4.88 \times 10^{43} \nonumber\\
&\approx& 2.74\times 10^{27} {\rm Wb}
 \approx (6.34\ {\rm gigavolts}) t_0,
\label{33}
\end{eqnarray}
where $t_0 \approx 13.7\, {\rm Ga}$ is the current age of the universe, or
\begin{equation}
N_{\ast} = 2eP_{\ast} = {2e^2\over m^2} \approx 8.33\times 10^{42},
\label{33b}
\end{equation}
such that an extreme classical black hole of this magnetic charge
would have a marginally strong magnetic field at its horizon,
\begin{equation}
b_{+} = {eB_{+}\over m^2} ={eP\over m^2 r_{+}^2}
 \approx {e\over m^2 P} = {e\over m^2 P_{\ast}} = 1.
\label{34}
\end{equation}

Now let
\begin{eqnarray}
q \equiv {P_{\ast}\over P} = {N_{\ast}\over N},
\quad u \equiv {P\over r},
\label{36}
\end{eqnarray}
so
\begin{equation}
b = {eB\over m^2} = q u^2.
\label{36b}
\end{equation}

The quantity $q$ is a constant for each extreme magnetic black hole
that is a measure of the strength of the field at the horizon.  It
take the value unity, $q=1$, for
\begin{equation}
M = M_{\ast} \approx P_{\ast} \approx 0.534\times 10^6 M_{\odot}
  \approx 788\,000\, {\rm km} \approx 2.63\, {\rm s}.
\label{37}
\end{equation}
Note that {\it smaller} extreme holes have {\it stronger} magnetic
fields at their horizons; $M < M_{\ast} \Rightarrow b_{+} \approx q > 1$.

The quantity $u$ is an inverse radial variable that goes from $u=0$ at
radial infinity to $u_{+} \approx 1$ at the horizon.  (One would have
the value of $u$ at the horizon, $u_{+}$, exactly 1 for the classical
Reissner-Nordstrom extreme black hole, but when the vacuum
polarization is taken into account, $u_{+}$ is shifted slightly away
from 1, as we shall see.)

Let
\begin{eqnarray}
g(b) &=& g(qu^2) = {e^2\over\pi}I(b),
\label{38}\\
f(b) &=& f(qu^2) = 1-g(b),
\label{39}
\end{eqnarray}
so
\begin{equation}
\rho = {B^2\over 8\pi}(1-g) = {B^2\over 8\pi}f.
\label{40}
\end{equation}

Now define the classically dimensionless mass function
\begin{equation}
\mu(u) \equiv {m(r=P/u)\over P}.
\label{41}
\end{equation}
Then the Einstein equation
\begin{eqnarray}
{dm\over dr} &=& 4\pi r^2 \rho = {1\over 2} r^2 B^2 [1-g(b)]
             = {1\over 2}{P^2\over r^2}(1-g)
\nonumber\\&=& {1\over 2} u^2(1-g)
             = {1\over 2} u^2 f
\label{42}
\end{eqnarray}
becomes
\begin{equation}
{d\mu\over du} = -{1\over 2} [1-g(qu^2)] = -{1\over 2} f(qu^2),
\label{43}
\end{equation}
where $f = 1-g(qu^2)$.

The boundary condition at radial infinity is that $m=M$ there, so
\begin{equation}
\mathcal{E} \equiv {M\over P} = {m_{\infty}\over P} = \mu_{\infty}
= \mu(u=0).
\label{45}
\end{equation}
At the horizon, $V \equiv 1-2m/r = 1-2\mu u = 0$, so the value of
$\mu$ at the horizon, $\mu_{+}$, is
\begin{equation}
\mu_{+} = {1\over 2u_{+}},
\label{46}
\end{equation}
always using the subscript $+$ to denote the value of a quantity at
the horizon.

For an extreme black hole, $dV/dr=0$ or $dV/du=0$ at the horizon,
which gives the equation
\begin{eqnarray}
{dV\over du} &=& -2\mu - 2u{d\mu\over du}
= -2\mu_{+} + u_{+}f(qu_{+}^2) \nonumber\\
&=& -{1\over u_{+}} + u_{+}f_{+} = 0,
\label{47}
\end{eqnarray}
or
\begin{equation}
u_{+} = [1-g(qu_{+})]^{-1/2},
\label{48}
\end{equation}
a parametric equation determining $U_{+} = u_{+}(q)$.

Then one integrates Eq. (\ref{43}) from $u=u_{+}$ to $u=0$ to
determine $\mathcal{E} = \mu(u=0)$:
\begin{eqnarray}
\mathcal{E} &\equiv& {M_{\rm extreme}\over P} = \mu(u=0)
   = \mu_{+} + \int_{u_{+}}^0 {d\mu\over du} du \nonumber \\
 &=& \mu_{+} + {1\over 2} \int_0^{u_{+}} du [1-g(qu^2)]
\nonumber\\
   &=& {1\over 2}\left[{1\over u_{+}}+u_{+}
       -\int_0^{u_{+}} du \, g(qu^2) \right]  \nonumber \\
 &=& {1\over 2}\left[f_{+}^{1/2}+f_{+}^{-1/2}
       -\int_0^{f_{+}^{-1/2}} du \, g(qu^2)\right],
\label{49}\\
f_{+} &=& 1-g(qu_{+}^2) = 1-g(qf_{+}^{-1}) = f_{+}(q) \nonumber \\
      &\approx& 1-g(q) = 1 - {\alpha\over\pi}I(q).
\label{50}
\end{eqnarray}

One can see that we have $g(q) = O(\alpha)$ and hence
$f_{+}^{1/2}+f_{+}^{-1/2} = 1+O(\alpha^2)$.  Therefore, to first order
in $\alpha = e^2$ (all that is given accurately by the one-loop
Euler-Heisenberg Lagrangian), we have
\begin{eqnarray}
\mathcal{E} &=& 1-{1\over 2}\int_0^1 g(qu^2) du + O(\alpha^2) \nonumber\\
 &\approx& 1-{\alpha\over 2\pi}\int_0^1 I(qu^2) du
 \equiv 1 - {\alpha\over 2\pi}J(q),
\label{53}
\end{eqnarray}
\begin{eqnarray}
{2\pi\over\alpha}\left(1-{M\over P}\right)+O(\alpha) = J(q) \equiv
 \int_0^1 du I(qu^2) \nonumber\\
=\int_0^{\infty} dx F(x){1\over 2}\sqrt{x\over q}
 \Gamma\left(-{1\over 2},{x\over q}\right).
\label{54}
\end{eqnarray}

Therefore, the mass-to-charge ratio $\mathcal{E} = M/P$ for an
extremal magnetically charged black hole is reduced from 1 (its
classical value for the Reissner-Nordstrom magnetic black hole without
any vacuum polarization) by an amount that to lowest order in the fine
structure constant $\alpha$ is $\alpha/(2\pi)$ times the quantity
$J(q)$ which depends on $q = P_{\ast}/P$, the ratio of the critical
magnetic charge to that of the actual magnetic charge.

Now we are left with the problem of estimating $J(q)$ for various
values of $q$, which can range from arbitrarily small values, for
arbitrarily large extremal black holes, to the huge value of $N_{\ast}
\sim 10^{43}$ for the smallest extremal magnetic charged black hole,
with $N=1$ and $P=1/(2e)\approx 5.853$ Planck units.

To use Eq. (\ref{54}) to calculate $J(q)$, it is useful to note that
the function $F(x)$ appearing therein and defined by Eq. (\ref{17})
may alternatively be written as
\begin{equation}
F(x) = {2x\over\pi^2}\sum_{k=1}^{\infty}{1\over k^2(\pi^2 k^2 + x^2)}.
\label{55}
\end{equation}
Then we get
\begin{eqnarray}
I(b) &=& \sum_{k=1}^{\infty}{1\over \pi^2 k^2}
       \int_0^{\infty} dw
\exp{\left(-{\pi k\over b}\sqrt{e^w-1}\right)} \nonumber \\
&=& {2\over \pi^2}\sum_{k=1}^{\infty}{1\over k^2}\Biggl[
-{\rm ci}\left({\pi k\over b}\right)\cos\left({\pi k\over b}\right)
-{\rm si}\left({\pi k\over b}\right)\sin\left({\pi k\over b}\right)\Biggr]
        \nonumber \\
&=& {1\over 3}[\ln(2b)-\gamma]  + {1\over 3}\sum_{n=1}^{\infty}
    \left({1\over n}-{1\over n+{1\over 2b}}\right)
\nonumber\\&&  -\sum_{m=1}^{\infty}{2\over (m+1)(m+2)}
\sum_{n=1}^{\infty}\left(n+{1\over 2b}\right)^{-m}
  \nonumber \\
&=& {1\over 3}\ln b -{1\over 3}\kappa
 \nonumber\\
&& +\sum_{m=1}^{\infty}{2\over (m+1)(m+2)}
   \sum_{n=1}^{\infty}\left({1\over n^m}
     -{1\over \left(n+{1\over 2b}\right)^m}\right),
\label{56}
\end{eqnarray}
where
\begin{eqnarray*}
\kappa = \gamma +\ln\pi-{\zeta'(2)\over \zeta(2)}
       \approx 2.291\,906\,543\,845\,465\,841\,149\,803\,801.
\label{57}
\end{eqnarray*}

From these formulas, we can get asymptotic expressions for $I(b)$ for
$b$ very small or very large.  For $b \ll 1$, we get the divergent
asymptotic series
\begin{eqnarray}
I(b) \sim -\sum_{n=1}^{\infty}{B_{2n+2}(2b)^{2n}\over n(n+1)(2n+1)}
     ={b^2\over 45}\Biggl(1-{4\over 7}b^2+{8\over 7}b^4
\nonumber\\
-{160\over 33}b^6
       +{176\,896\over 5005}b^8-{5120\over 13}b^{10} +O(b^{12})\Biggr).
\label{58}
\end{eqnarray}
For $b \gg 1$, we get
\begin{eqnarray}
I(b) &\sim& \left({1\over 3}+{1\over b}+{1\over 2b^2}\right)\ln b
  -{1\over 3}\kappa + {2-\ln\pi\over b}
  \nonumber\\&& +{3-2\gamma+2\ln 2\over 4b^2} + {\pi^2\over 72b^3}
  + O\left({1\over b^4}\right).
\label{59}
\end{eqnarray}

We can also write various expressions for $J(q)$, going beyond
Eq. (\ref{54}), such as
\begin{eqnarray}
J(q) &=& \int_0^1 du\int_0^{\infty} dx
   \left({1\over x^3}+{1\over 3x}-{\coth x\over x^2}\right)
    \exp{\left(-{x\over qu^2}\right)}  \nonumber \\
&=&\sum_{k=1}^{\infty}{1\over \pi^2 k^2}
    \int_0^{\infty}{z\, dz\, \exp{(-z)}\over z^2+\pi^2 k^2/q^2}
     \sum_{n=1}^{\infty}{L_n^{-1/2}(z)\over n+1} \nonumber \\
&=&\sum_{k=1}^{\infty}{4\over \pi^2 k^2}
    \int_0^{\infty}{x^3 dx\over x^4 +\pi^2 k^2/q^2}
     \left[\exp{\left(-{x^2}\right)}
         -\sqrt{\pi} x\, {\rm erfc}\, x \right] \nonumber \\
&=&\left({1\over 3}-{1\over q}-{1\over 6q^2}\right)\ln q
   -{1\over 3}(\kappa+2)
+\sqrt{2}\,\zeta\left({3\over 2}\right){1\over\sqrt{q}} \nonumber\\&&
   -{2+4\ln 2 + 3\ln\pi\over q}
+{6\gamma-13-6\ln 2\over 36q^2}
 \nonumber\\&&  +4\sum_{k=2}^{\infty}{\zeta(k)\over k(k+1)(2k+1)}
      \left(-{1\over 2q}\right)^{k+1}.
\label{60}
\end{eqnarray}
The last expression for $J(q)$, with the one infinite series and no
integrals, converges for $q > 1/2$.

Now we can also get asymptotic formulas for $J(q)$ for both $q$ very
small and for $q$ very large.  For $q \ll 1$, we get the divergent
asymptotic series
\begin{eqnarray}
J(q) &\sim& {q^2\over 225}\Biggl(1-{20\over 63}q^2+{40\over 91}q^4-{800\over 561}q^6
 \nonumber\\&& +{176\,896\over 21\,021}q^8-{1024\over 13}q^{10} +O(q^{12})\Biggr).
\label{61}
\end{eqnarray}
For $q \gg 1$, inserting numerical values for the numerical
coefficients of the terms before the series in the last expression of
Eq. (\ref{60}) gives
\begin{eqnarray}
J(q) &\sim& {1\over 3}\ln q -1.43063551 + {3.69445665\over\sqrt{q}}
      \nonumber\\&& -{\ln q + 8.20677838\over q} - {\ln q\over 6q^2}
      + O\left({1\over q^3}\right).
\label{61b}
\end{eqnarray}

A simple fitting function that matches the two leading terms at both
$q \ll 1$ and $q \gg 1$ is
\begin{eqnarray}
\tilde{J}(q)={1\over 6}[\!\!&+&\!\!0.9583958466\ln(1+0.0001314447009 q^2)
      \nonumber \\
                        &+&\!\!0.0416041534\ln(1+0.6379336776 q^2)].
\label{62}
\end{eqnarray}
However, this drops down to $0.30608 J(q)$ at $q=26.355$.  One can
get a fit to within 5-6\% accuracy for all $q$ (and fitting with
arbitrarily high accuracy for arbitrarily large or small $q$) by
multiplying $\tilde{J}(q)$ by the exponential of a suitable gaussian
in $\ln q$:
\begin{equation}
J(q) \approx \hat{J}(q) =
 \tilde{J}(q)\exp{\{1.185\exp{[-0.225(\ln q - \ln 26)^2]}\}}.
\label{63}
\end{equation}

Remember that $J(q)$ is given precisely by the double integrals of
Eqs. (\ref{54}) and (\ref{60}), and that to lowest nontrivial order in
the fine structure constant $\alpha$, the effect of the vacuum
polarization of the electron-positron field of charge $e$ and mass $m$
(and ignoring the vacuum polarization effects of other fields, which
are smaller, at least for sufficiently large black holes that $q$ is
not too much larger than unity) gives, by Eqs. (\ref{11}), (\ref{53}),
(\ref{62}), and (\ref{63}),
\begin{equation}
\mathcal{E} \equiv {M_{\rm extreme}\over P}
 \approx 1 - {\alpha\over 2\pi}J(q)
 \approx 1 - {\alpha\over 2\pi}\hat{J}(q),
\label{64}
\end{equation}
where Eqs. (\ref{33}), (\ref{36}), and (\ref{37}) give
\begin{eqnarray}
q &\equiv& {P_{\ast}\over P} = {e\over m^2 P}
  = {4.88\times 10^{43}\over P}
  = {2.74\times 10^{27}{\rm Wb}\over P} \nonumber \\
 &=& {0.534\times 10^6 M_{\odot}\over P}
  = {7.88\times 10^8 \, {\rm m}\over P} = {2.63 \, {\rm s}\over P},
\label{65}
\end{eqnarray}
with $P$ being the magnetic charge (and hence approximately the mass
$M_{\rm extreme}$) of the extreme magnetically charged black hole.

These formulas apply only for $P$ large in Planck units, $P \gg 1$.
Since these formulas also assume that the vacuum polarization of the
electron-positron field dominates, they actually apply with good
accuracy only when the analogous $q$'s for heavier charged fields,
such as the muon, are small.  Taking these fields to have masses at
least two orders of magnitude larger than the electron-positron field,
so that the corresponding $q$'s are at least four orders of magnitude
smaller than that for the electron-positron field, means that the
formulas above should be good so long as the $q$ for the
electron-positron field is much smaller than about $10^4$, or $P \gg
100 M_{\odot}$.

\section{Small extreme magnetic black holes}

Let us now go to the other extreme, where $q \gg 10^4$ or $P \ll 100
M_{\odot}$.  First consider the case of the smallest extreme magnetic
black holes.

The minimum value for the magnetic charge $P = N/(2e)$ is $P=1/(2e)$,
for $N=1$, giving $q = P_{\ast}/P = 2e^2/m^2 = 8.33\times 10^{42}$.
Then
\begin{eqnarray}
J(q) \approx {1\over 3} \ln q
 - {1\over 3}(\gamma+\ln\pi-{\zeta'(2)\over\zeta(2)}+2)
 \approx 31.5123,
\label{66}
\end{eqnarray}
so
\begin{eqnarray}
\delta(P) = 1 - {M\over P} \approx {\alpha\over 2\pi}J(q)
\approx 0.0365987 \approx {1\over 27.3234},
\label{67}
\end{eqnarray}
giving
\begin{equation}
\mathcal{E} = {M\over P} = 1 - \delta(P) \approx 0.9634013.
\label{68}
\end{equation}

However, this just includes the vacuum polarization effects of the
electron-positron field.  The one-loop effect of all charged Dirac
fields of charge magnitude $e_i$ and mass $m_i \ll 1$ is
\begin{eqnarray*}
1-\mathcal{E} = \delta
 \approx {1\over 6\pi}\sum_i e_i^2
   \left[\ln{2ee_i\over m_i^2}-\gamma-\ln\pi+{\zeta'(2)\over\zeta(2)}-2\right].
\label{69}
\end{eqnarray*}

Including all three charged leptons (electrons, muons, and taus) gives
\begin{eqnarray*}
J_{\rm leptons} \approx 85.55,\ \ \delta_{\rm leptons} \approx 0.09935,\ \
1-\delta_{\rm leptons} \approx 0.90065.
\label{70}
\end{eqnarray*}
Including three colors of the six quark flavors with masses in GeV
taken to be (0.003, 0.006, 0.123, 1.25, 4.2, 171) gives
\begin{eqnarray*}
J_{\rm quarks} \approx 133.07,\ \ \delta_{\rm quarks} \approx 0.15455,\ \
1-\delta_{\rm quarks} \approx 0.84545.
\label{71}
\end{eqnarray*}
Including all of these charged fields gives
\begin{eqnarray*}
J \approx 218.62,\ \ \delta \approx 0.2539,\ \
\mathcal{E} = 1-\delta \approx 0.7461 \approx {M\over P}.
\label{72}
\end{eqnarray*}
Thus if these numerical values are correct and give the dominant
vacuum polarization, the smallest extreme magnetic black hole might
have a mass that is about $25\%$ less than what the classical
Reissner-Nordstrom metric would indicate without vacuum polarization.

Now let us go to somewhat larger extremal black holes, but such that
the $q$'s for all of the lepton and quark fields are large.  That is,
we shall now consider any $N = 2eP \ll e^2/m_{\rm top}^2 \sim 4\times
10^{31}$. Then
\begin{eqnarray}
\mathcal{E} &=& {M\over P}
= 1 - \delta \nonumber\\ &\approx& 1 - {1\over 6\pi}\sum_i e_i^2
   \Biggl[\ln{2ee_i\over
   m_i^2}-\gamma -\ln\pi+{\zeta'(2)\over\zeta(2)}-2\Biggr]
     \nonumber\\
 &\sim& 0.75 + 0.0031 \ln N,
\label{73}
\end{eqnarray}
assuming vacuum polarization purely from quarks and leptons.  Thus
extreme magnetic black holes with $P \ll e/m_{\rm top}^2 \sim 4\times
10^{32} \sim 10^{25} \, {\rm kg} \sim 1\, {\rm cm}$ might have
\begin{eqnarray}
M = M(P) &\sim& 0.75 P + 0.0031 P \ln P  \nonumber \\
         &\approx& (0.75 + 0.0031 \ln N){N\over 2e}  \nonumber \\
         &\approx& (4.4 + 0.018 \ln N)N.
\label{74}
\end{eqnarray}

If an extreme black hole of magnetic charge $P = N/(2e)$ splits into
two extreme holes of charges $P_1 = N_1/(2e)$, $P_2 = N_2/(2e)$, with
$N_1 + N_2 = N$, then the energy released into kinetic energy is
\begin{eqnarray}
\Delta E &=& M(P)-M(P_1)-M(P_2) \nonumber \\
 &\approx& {2e\over 3\pi}(N\ln N - N_1\ln N_1 - N_2\ln N_2)
 \nonumber \\
 &\approx& 0.018\left(N_1\ln{N\over N_1}+N_2\ln{N\over N_2}\right).
\label{75}
\end{eqnarray}
If $N_1 = N-1 \gg 1$, $N_2 = 1$,
\begin{equation}
\Delta E \approx {2e\over 3\pi}(1 + \ln N).
\label{76}
\end{equation}

\section{Renormalization group estimate}

Is it a coincidence that $1-\mathcal{E} = \delta = O(1)$?
By a crude renormalization group calculation using the Minimal
Supersymmetric Standard Model and the approximation $-\ln m_{\rm
proton} \sim -\ln m_{\rm Higgs} \sim -\ln m_{\rm SUSY} \gg -\ln m_{\rm
GUT} \sim -\ln m_{\rm Pl} \equiv 0$, one of us showed \cite{Pag} that
\begin{eqnarray*}
-\ln m_{\rm proton} \sim {\pi\over 10 e^2}.
\label{77}
\end{eqnarray*}

Then if we have $n_l = 3$ leptons with $-\ln m_l$ comparable to $-\ln
m_{\rm proton}$, taking only the leading terms in Eq. (\ref{69}) gives
\begin{eqnarray*}
\delta_{\rm lepton} &\approx& {e^2\over 2\pi}J
 \sim {e^2\over 2\pi}{n_l\over 3}\ln{2e^2\over m_l^2}
 \sim {e^2\over 2\pi}{n_l\over 3}(-2\ln m_l) \nonumber \\
 &\sim& {e^2\over 2\pi}{n_l\over 3}(-2\ln m_{\rm proton})
 \sim {e^2\over 2\pi}{n_l\over 3}{\pi\over 5e^2}
 = {n_l\over 30} = {1\over 10},
\label{78}
\end{eqnarray*}
or $1-\delta_{\rm lepton} \sim 0.9$, which is accidentally extremely
close to the value calculated above, 0.90065.
For quarks and leptons all of similar $-\ln m_i$, one gets $\delta
\sim 4/15$, $1-\delta \sim 11/15 \approx 0.73$ instead of the value
$1-\delta \approx 0.75$ estimated above.

Thus it seems to be no accident that $1-\mathcal{E} = \delta = O(1)$,
and indeed the value depends mainly on the number of species of
charged particles and is rather insensitive to the actual value of the
charge $e$ or the fine structure constant $\alpha$.

\section{Decay rates for extreme black holes}

What is the decay time for an extreme magnetically charged black hole
of magnetic charge $P = N/(2e) \gg 1/(2e)$ and mass $M \approx P$ to
emit a minimal extremal hole of $P=g=1/(2e)$ and mass
$m=\mathcal{E}g$?

Ignoring prefactors, the time is $t \sim e^{2I}$ with tunneling action
\begin{equation}
I = \int_{r_1}^{r_2} \sqrt{-p_r^2}\, dr,
\label{79}
\end{equation}
where the radial momentum $p_r$ is given by
\begin{eqnarray*}
0 = g^{\alpha\beta}\pi_{\alpha}\pi_{\beta}+m^2
  = -V^{-1}\left(E-{gP\over r}\right)^2 + V p_r^2 + m^2,
\label{80}
\end{eqnarray*}
\begin{eqnarray*}
V \approx \left(1-{P\over r}\right)^2, \quad m \leq E \leq gP/r_{+} \approx g.
\label{81}
\end{eqnarray*}
The minimum action $I$ is for $E=g=m/\mathcal{E} > m$:
\begin{eqnarray}
2I_{\rm min} &=& 2\pi gP(1-\sqrt{1-\mathcal{E}^2}) \nonumber\\
 &=& {\pi P\over e}(1-\sqrt{1-\mathcal{E}^2})
 = {\pi N\over 2\alpha}(1-\sqrt{1-\mathcal{E}^2}) \nonumber \\
&\approx& 36.78 P (1-\sqrt{1-\mathcal{E}^2})
 \approx 215.26 N (1-\sqrt{1-\mathcal{E}^2}) \nonumber \\
&\sim& 12.33 P \sim 72.18 N,
\label{82}
\end{eqnarray}
with $\mathcal{E} \sim 0.7461$ to get the numerical results on the
last line.

The time for the extremal magnetic black hole to decay is then $t \sim
e^{2I_{\rm min}} \sim e^{72.18N} \approx 10^{31.35N}$.
For $P=M_{\odot}=9.137\times 10^{37}$ or $N=2eP=1.561\times 10^{37}$,
\begin{equation}
t \sim e^{1.13\times 10^{39}} \approx 10^{4.89\times 10^{38}}
 \approx 10^{10^{38.69}}.
\label{83}
\end{equation}
This is much, much greater than a googol, but much, much less than
googolplex.

If the extreme magnetic black hole emitted magnetic monopoles of
$\mathcal{E} = m/g \ll 1$ instead, one would get $t \sim e^{\pi g P
\mathcal{E}^2} = \exp{\left({\pi m^2\over g B_{+}}\right)}$, the inverse
Schwinger rate for the field strength at the horizon.  For a general
value of $\mathcal{E}$, one would get  $t \sim e^{2I_{\rm min}}$ with
\begin{equation}
2I_{\rm min} = {2\over 1+\sqrt{1-\mathcal{E}^2}}{\pi m^2\over g B_{+}}.
\label{84}
\end{equation}

\section{Energy and entropy of near extreme black holes}

From the results of the previous section, near-extreme magnetic holes
take $t \sim e^{2I} = e^{aN}$ to emit a minimal hole, with
\begin{equation}
a = {\pi\over 2\alpha}(1-\sqrt{1-\mathcal{E}^2}) \approx 72.18.
\label{85}
\end{equation}

If we use the results of \cite{Page} under the preferred assumption
there that no energy eigenstates of a black hole have high degeneracy,
in the time it takes a near-extreme magnetic black hole to lose
another unit of its magnetic charge, by photon emission it would get
down to having excess energy
\begin{eqnarray*}
E \equiv M - M_{\rm extreme}(P) \sim N^{-{1\over 13}}t^{-{2\over 13}}
 \sim e^{-11.11N}.
\label{86}
\end{eqnarray*}

Since this gives $E \ll P^{-3}$, if the hole is thermalized, it has
temperature $T \approx E \sim e^{-11.11N}$ and entropy
\begin{eqnarray*}
S&\approx& {1\over 4}A+\ln{(2 \pi^2 P^3 E)}
 \approx {1\over 4}A - 11.11N + {38\over 13}\ln N \nonumber \\
 &\approx& {\pi\over 4e^2} N^2 - 11.11N
 \approx 107.62782253 N^2 - 11.11N.
\label{87}
\end{eqnarray*}

For $P=M_{\odot}=9.137\times 10^{37}$, $N=2eP = 1.561\times 10^{37}$,
one gets
\begin{equation}
T \sim e^{-1.734\times 10^{38}}
  \approx 10^{-7.529\times 10^{37}}
  \approx 10^{-10^{37.8767}},
\label{88}
\end{equation}
\begin{equation}
S \sim {1\over 4}A-1.734\times 10^{38}
 \approx 2.623\times 10^{76} - 1.734\times 10^{38}.
\label{89}
\end{equation}

\section{Open questions for future work}

(1)  What is the effect of the vacuum polarization from all charged
particle fields, and not just that from leptons and quarks, on $M(P)$?

(2)  What is the effect of the Nielsen-Olesen-Skalozub phase
transition \cite{NO,Sk} for $B > m_W^2/e$?  This arises from the fact
that the charged vector boson $W$ develops a negative-energy
Landau level in the na\"{\i}ve vacuum and thereafter gives an imaginary
contribution to the Euler-Heisenberg Lagrangian for the magnetic field
in the na\"{\i}ve vacuum, signifying its instability.

(3) Since the one-loop effects are not very small, what are the
effects of higher loops?

\section{Conclusions}

Vacuum polarization effects of charged particle fields, in the
presence of the magnetic field of an extreme magnetically charged
black hole, can reduce the mass of the hole below that given by the
classical extreme Reissner-Nordstrom metric with no vacuum
polarization.  Since the reduction in the mass is greater for smaller
extreme black holes, which have stronger magnetic fields at their
horizons, this makes it energetically favorable for larger extreme
magnetic black holes to split up into smaller ones.  Thus magnetic
black holes are unstable to splitting, even if there are no
non-black-hole magnetic monopoles that they can decay into.

In asymptotically flat spacetime, the entropy of the kinetic energy
released in the process can exceed the entropy decrease of the holes
themselves and make the process thermodynamically allowed.  However,
for large extremal black holes, there is a very large action barrier,
so the decay process is extremely slow.  During the decay process,
there is plenty of time for the temperature to drop to exponentially
tiny values (assuming no incoming radiation), and the entropy of the
hole can drop below $A/4$ by an absolute amount that is very large,
even though still extremely tiny in comparison with the much greater
value of $A/4$.

\acknowledgments

The work of S.P.K. was supported by the Korea Research Foundation under Grant
No. 1999-2-112-003-5, and
the work of D.N.P. was supported in part by the Natural Sciences and
Engineering Research Council of Canada.

\end{document}